\documentclass[aps, prd, onecolumn, amssymb, amsmath, showpacs, superscriptaddress]{revtex4-1}
\usepackage[T1]{fontenc}
\usepackage[latin9]{inputenc}
\usepackage{color}
\usepackage{float}
\usepackage{amsmath}
\usepackage{graphicx}
\usepackage{amssymb}
\usepackage{esint}
\usepackage{SIunits}

\begin{document}

\title{Floquet High Chern Insulators in Periodically Driven Chirally Stacked Multilayer Graphene}

\author{Si Li}
\affiliation{Beijing Key Laboratory of Nanophotonics and Ultrafine Optoelectronic Systems, School of Physics, Beijing Institute of Technology, Beijing 100081, China}

\author{Cheng-Cheng Liu}
\email{ccliu@bit.edu.cn}
\affiliation{Beijing Key Laboratory of Nanophotonics and Ultrafine Optoelectronic Systems, School of Physics, Beijing Institute of Technology, Beijing 100081, China}

\author{Yugui Yao}
\affiliation{Beijing Key Laboratory of Nanophotonics and Ultrafine Optoelectronic Systems, School of Physics, Beijing Institute of Technology, Beijing 100081, China}

\begin{abstract}
Chirally stacked $N$-layer graphene is a semimetal with $\pm p^N$ band-touching at two nonequivalent corners in its Brillioun zone. We predict that an off-resonant circularly polarized light (CPL) drives chirally stacked $N$-layer graphene into a Floquet Chern Insulators (FCIs), a.k.a. quantum anomalous Hall insulators, with tunable high Chern number $C_F=\pm N$ and large gaps. A topological phase transition between such a FCI and a valley Hall (VH) insulator with high valley Chern number  $C_v=\pm N$ induced by a voltage gate can be engineered by the parameters of the CPL and voltage gate. We propose a topological domain wall between the FCI and VH phases, along which perfectly valley-polarized $N$-channel edge states propagate unidirectionally without backscattering.
\end{abstract}
%\pacs{73.43.-f, 81.05.ue}
\maketitle
%\noindent{\bf Introduction}

\maketitle

\section{Introduction}
The rise of graphene has triggered tremendous efforts to investigate the novel physical properties of its multilayer versions both experimentally and theoretically~\cite{Neto2009,Ohta2006,Zhang2009,Lui2011,McCann2006,Min2008,Zhang2010,Jung2011}. Although these multilayers bound in monolayer graphene by a weak Van der Waals force, their low-energy spectra are fundamentally different from that of monolayer one and sensitive to the number of layers and stacking orders. For the chirally (ABC) stacked $N$-layers, also referred to as chiral 2DEGs~\cite{Min2008}, the spectra consist of two bands touching at two valleys $K$ and $K'$ and the other $2N-2$ bands located at higher energies. Properties of quasiparticle excitations in chiral 2DEGs are determined by their chirality index $N$. The low energy conduction and valence bands in chirally (ABC) stacked $N$-layer graphene have $\pm p^N$ dispersion (here $p$ is momentum measured from the two valleys $K$ and $K'$) and a Berry phase of $N\pi$, $N$ times the value of Dirac fermions. These unique properties could result in remarkable features such as unusual Landau quantization.

Recently, there has been broad interest in the condensed matter physics community in the search for novel topological phases, aiming for both scientific explorations and potential applications~\cite{Hasan2010,Qi2011}. Among these novel topological phases is the Chern insulator, also known as the quantum anomalous Hall insulator~\cite{Haldane1988,Onoda2003,Liu2008,Yu2010,Wangzf2013,Xu2015,Jiang2012,Trescher2012,Yang2012}, which is characterized by a non-zero Chern number $C$~\cite{Thouless1982}, and maintains robust stability against disorder and other perturbations. Although the first proposal appeared over twenty years ago, not until recently was the experimental evidence for the Chern insulator phase with $C=1$ reported within Cr-doped (Bi,Sb)$_2$Te$_3$ at extremely low temperatures~\cite{Chang2013,Kou2014}. Some strategies to achieve Chern insulators at high temperatures in honeycomb materials are proposed, such as by transition-metal atoms adaption~\cite{Qiao2010,Ding2011,Zhang2012,Zhang2013}, magnetic substrate proximity effect~\cite{Garrity2013,Qiao2014,Ezawa2012,Pan2014}, and surface functionalization~\cite{Huang2014,Wu2014,Liu2015,Niu2015}. In honeycomb lattices, $K$ and $K'$ valleys provide a tunable binary degree of freedom to design valleytronics. By breaking inversion symmetry, a bulk band gap can be opened to host a valley Hall (VH) effect, which is classified by a valley Chern number $C_v = C_K-C_{K'}$ ~\cite{Xiao2007,Martin2008,Gorbachev2014}. For general technological applications, it is urgent and important to find these topological states of high quality  such as with high tunable topological number and a large gap.

\begin{figure}[t!]
\includegraphics[width=0.8\columnwidth]{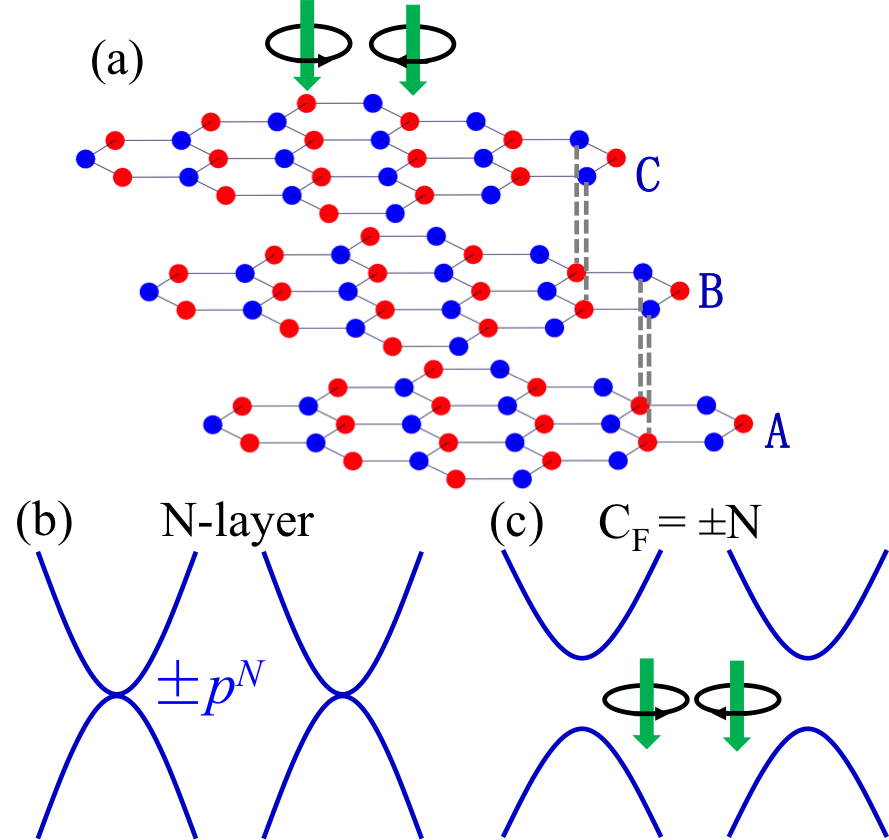}
\caption{Schematics of creating Floquet Chern insualors (FCIs) with high Chern number by irradiating an off-resonant left/right circularly polarized light (CPL) on chirally (ABC) stacked multilayer graphene. (a) Illuminating the Left/right CPL on the  multilayer graphene. (b) Energy dispersion for free-standing chirally stacked $N$-layer graphene. (c) A left/right CPL induces a mass term and drive the system to FCI phases with Chern number $C_F=\pm N$.}
\label{fig1}
\end{figure}

Light irradiation by a Laser maybe an ideal means to engineer the band structures in a high frequency limit~\cite{Cayssol2013,Bukov2015,Rechtsman2013,Jotzu2014}, taking into account that a Laser has many merits, e.g. monochromaticity, strength and energy can be adjusted in a large range, and without introducing impurities. Periodically driven quantum systems by an off-resonant circularly polarized light (CPL) can be described in the Floquet theory framework. A CPL naturally breaks time-reversal symmetry, which may lead to exotic topological phases. Shining light radiation by an off-resonant CPL on a trivial insulator~\cite{Inoue2010,Lindner2011,Dora2012}, monolayer graphene~\cite{Oka2009,Kitagawa2011,Zhai2014,Sentef2015,Leon2014,Piskunow2014}, silicene~\cite{Ezawa2013}, nodal line semimetals ~\cite{Yan2016,Chan2016,Taguchi2016,Saha2016} and so forth could induce distinguishing topological phases. In addition, compared with the static systems, the Floquet edge modes can appear in the edge of the driven system even though all the topological invariants (Chern numbers) of the bulk bands are  trivial~\cite{Rudner2013, Yao2017}. The Floquet-Bloch states have been experimentally observed on the surface of photo-excited Bi$_2$Se$_3$ by using time- and angle-resolved photoemission spectroscopy~\cite{Wang2013,Fregoso2013}.

Such topologically nontrivial states with high topological number in controllable manners and a large gap are quite scarce in nature due to their extremely stringent requirements on special band structures and sample quality.
In this paper, we demonstrate that periodically driven chirally stacked finite $N$-layer graphene systems by a left/right CPL can become Floquet Chern insualors (FCIs) with tunable high Chern number $\pm N$ just by number of the layers, and a large gap. In addition, a gate voltage can also drive the chirally stacked $N$-layer graphene into VH phases with high valley Chern number $\pm N$ by taking advantage of the feature that its low-energy bands come from the respective top layer and bottom layer. Using the two kinds of FCI and VH insulators, we design a topological domain (DW) wall, along which $N$-channel perfect valley-polarized chiral edge states emerge. Hopefully, our finds will motivate further investigations on the fascinating properties of periodically driven chirally stacked $N$-layer graphene systems from both basic and applied research fields.

\section{Model and method}
The tight-binding Hamiltonian for the chirally stacked $N$-layer graphene systems with only perpendicular hopping as interlayer interaction reads
\begin{equation}\label{H0TB}
H_0=\sum_{n=1}^{N}h_{0n}+\sum_{n=2}^{N}h_{\perp}+\sum_{n=1}^{N}h_{U}.
\end{equation}
The first term $h_{0n}=-t\sum_{\left\langle i,j\right\rangle }c_{in}^{\dagger}c_{jn}$ is the intralayer nearest neighbor hopping term, the second term $h_{\perp}=\sum_{\left\langle i,j\right\rangle }t_{\perp}\left(c_{in}^{\dagger}c_{jn-1}+h.c.\right)$ is the perpendicular interlayer hopping between B-sublattice in the $(n-1)^{th}$, $B_{n-1}$ and A-sublattice in the $n^{th}$, $A_n$, and the last one $h_{U}=\sum_{i}U_{n}c_{in}^{\dagger}c_{in}$ is staggered potential term with $U_{1}=U,U_{N}=-U,U_{n}=U-\frac{2\left(n-1\right)}{N-1}U$.
In the basis $\{A_1,B_1,A_2,B_2,...,A_N,B_N\}$, the $2N\times2N$ Hamiltonian near the valley $K$ and $K'$ is a banded matrix with width of three. $H_{0}\left(\boldsymbol{p}\right)_{i,i\mp1}=\left(v_{F}p_{\pm},t_{\bot},....,v_{F}p_{\pm}\right)_{2N-1}$ with $p_{\pm}=\tau_{z}p_x\pm ip_y$ and $H_{0}\left(\boldsymbol{p}\right)_{i,i}=\frac{2U}{N-1}\left(\frac{N-1}{2},\frac{N-1}{2},\frac{N-3}{2},\frac{N-3}{2},...,-\frac{N-1}{2},-\frac{N-1}{2}\right)_{2N}$, where the Fermi velocity $v_{F}=\sqrt{3}/2at\sim c/300$, lattice constant $a$=2.46 \AA, and $\tau_{z}=\pm1$ labels the $K$ and $K'$ valleys. Via the similar downfolding procedure~\cite{Min2008,Zhang2011}, the N-step process where electrons hop between low-energy sites on the two outermost layers ($A_1$ and $B_N$) through high-energy states can be described by a effective $2\times2$ band continuum model $H_{0,eff}\left(\boldsymbol{p}\right)=g_{N}\left[\left(\tau_{z}p_{x}+ip_{y}\right)^{N}\sigma_{-}+\left(\tau_{z}p_{x}-ip_{y}\right)^{N}\sigma_{+}\right]+U\sigma_{z}$, with $g_{N}\equiv\left(v_{F}\right)^{N}/\left(-t_{\perp}\right)^{N-1}$. The Pauli matrices  $\boldsymbol{\sigma}$ act on the low-energy layer pseudospin degree of freedom with $\sigma_{\pm}=\left(\sigma_{x}\pm\sigma_{y}\right)/2$.  Without the staggered potential term, the gapless dispersion is $\pm p^N$, as shown in Fig.~\ref{fig1}(b). The gapless dispersion is protected by time-reversal symmetry and inversion symmetry in spinless case. Taking into account the staggered potential term, the $N$-layer chirally stacked graphene becomes an insulator with a 2$\left|U\right|$ gap.

\section{Photoinduced Floquet Chern insulators with high Chern number}
We now study the effects of a periodic driving by applying a beam of off-resonant circularly polarized light (CPL) onto the $N$-layer chirally stacked graphene systems perpendicularly. The related vector potential $\boldsymbol{A}\left(t\right)=A_{0}\left(cos(\omega t),\xi sin(\omega t)\right)$ minimally couples to the systems via replacing the crystal momentum $\boldsymbol{p}$ by the covariant momentum $\boldsymbol{p}+e\boldsymbol{A}\left(t\right)$ in the above $H_{0}\left(\boldsymbol{p}\right)$, where $\xi=1\left(-1\right)$ denotes the right- (left-) CPL, and $\omega$ are the frequency of the light. The full $2N\times2N$ Hamiltonian $H\left(\boldsymbol{p},t\right)$ is time-periodic (with periodicity $T=2\pi/\omega$) in the presence of the periodic driving light field,
\begin{equation}\label{time-periodic}
H\left(\boldsymbol{p},t\right)=H_{0}\left(\boldsymbol{p}+e\boldsymbol{A}\left(t\right)\right)=H_{0}\left(\boldsymbol{p}\right)+H_{0}\left(e\boldsymbol{A}\left(t\right)\right).
\end{equation}
Since the elements of $H_{0}$ is linear in $\boldsymbol{p}$ at most, one has the second equation. Here, $H_{0}\left(e\boldsymbol{A}\left(t\right)\right)=I_{N}\otimes\left(\tau_{z}cos\left(\omega t\right)\sigma_{x}+\xi sin\left(\omega t\right)\sigma_{y}\right)eA_{0}v_{F}$, with $I_{N}$ the $N$-order identity matrix.

In the regime of Floquet picture, the high frequency limit is taken into account, where the frequency of the CPL is far greater than the other characteristic energy scales. Such so-called off-resonant light does not directly contribute to the optical transition of electrons between different energy levels, but results in photon-dressed band structures' modification via virtual photon emission/absorption processes. Here we take the smallest energy of the CPL as the band width $6t\simeq16$ eV, whose frequency is 2.4$\times$10$^{16}$ Hz (UV-light). The typical experimental field strength is given in a range $eA_0$=0.01-0.2 \AA$^{-1}$~\cite{Saha2016}, and the corresponding Laser intensity $I=\left(eA_0\omega\right)^2/8\pi\alpha\hbar$=0.34-136 $\times 10^{12}$ W/cm$^2$ with $\alpha$ being the fine structure constant. The effective Floquet Hamiltonian can then be expanded in a static form up to $1/\omega$ as follows (the details are shown in appendix A):
\begin{equation}\label{Floquet}
H_{F}\left(\boldsymbol{p}\right)=	H_{0}\left(\boldsymbol{p}\right)+H_{1\omega}\left(\boldsymbol{p}\right)+\mathcal{O}\left(\frac{1}{\omega^{2}}\right),
\end{equation}
where $H_{1\omega}\left(\boldsymbol{p}\right)=\frac{1}{\hbar\omega}\sum_{n=1}^{\infty}\frac{1}{n}\left[H_{n},H_{-n}\right]$, with $H_{n}=\frac{1}{T}\int_{0}^{T}H\left(\boldsymbol{p},t\right)e^{-in\omega t}dt$.
Here only $H_1$ and $H_{-1}$ are nonzero and expressed as
$H_{\pm1}=I_{N}\otimes\left(\tau_{z}\sigma_{x}\mp i\xi\sigma_{y}\right)\frac{eA_{0}v_{F}}{2}$.
The CPL induced part reads explicitly,
\begin{equation}\label{CPL-induced}
H_{1\omega}\left(\boldsymbol{p}\right)=I_{N}\otimes\left(-1\right)\frac{\left(eA_{0}v_{F}\right)^{2}}{\hbar\omega}\xi\tau_{z}\sigma_{z}.
\end{equation}
Due to its diagonal form, in the low-energy two basis $\{A_1,B_N\}$, the effective $2\times2$ Hamiltonian of Eq.~\ref{CPL-induced} reads $H_{1\omega ,eff}\left(\boldsymbol{p}\right)=-m_F\xi\tau_{z}\sigma_{z}$, with the Floquet mass $m_F\equiv\frac{\left(eA_{0}v_{F}\right)^{2}}{\hbar \omega}$. As a result, the effective $2\times2$ Floquet Hamiltonian for irradiated chirally stacked $N$-layer graphene by a CPL is expressed as
\begin{equation}\label{Floquet-eff}
H^F_{N,eff}\left(\boldsymbol{p}\right)=g_{N}\left[\left(\tau_{z}p_{x}+ip_{y}\right)^{N}\sigma_{-}+h.c.\right] +\lambda\sigma_{z},
\end{equation}
with $\lambda\equiv U-m_F\xi\tau_{z}$. The second term is a mass term added to the otherwise $p^N$ crossing gapless $N$-layer chirally stacked graphene. The spectrum for the Floquet effective Hamiltonian is $E=\pm\sqrt{g_{N}^{2}p^{2N}+\lambda^{2}}$ with a gap of  2$\left|\lambda\right|$. Besides a gate voltage, the sign and size of the mass at respective $K$ and $K'$ valleys can be tuned by the chirality, frequency, and strength of a CPL. When $N$=1, the photo-induced part is no other than the Haldane term~\cite{Kitagawa2011} for mono-layer graphene model~\cite{Haldane1988} in addition to the chirality dependence of CPL. We generalize the case of mono-layer to $N$-layers and develop a tight-binding model for the irradiated chirally stacked $N$-layer graphene,
\begin{equation}\label{H_FloquetTB}
\begin{split}
H_{F}&= \sum_{n=1}^{N}\left(-t\sum_{\left\langle i,j\right\rangle }c_{in}^{\dagger}c_{jn}+i\frac{t_{p}}{3\sqrt{3}}\sum_{\left\langle \left\langle i,j\right\rangle \right\rangle }\nu_{ij}c_{in}^{\dagger}c_{jn}\right)\\
&+\sum_{n=2}^{N}\left(t_{\perp}\left(c_{in}^{\dagger}c_{jn-1}+h.c.\right)\right)+\sum_{n=1}^{N}\left(U_{n}c_{in}^{\dagger}c_{in}\right),
\end{split}
\end{equation}
where the second new added term is the Haldane-like next-nearest neighbor hopping term induced by CPL with $t_{p}=\xi m_F$. The $\nu_{ij}=\pm1$ corresponds to the next-nearest neighbor hopping anticlockwise or clockwise with respect to the positive $z$ direction.

\begin{figure}[b!]
\includegraphics[width=0.9\columnwidth]{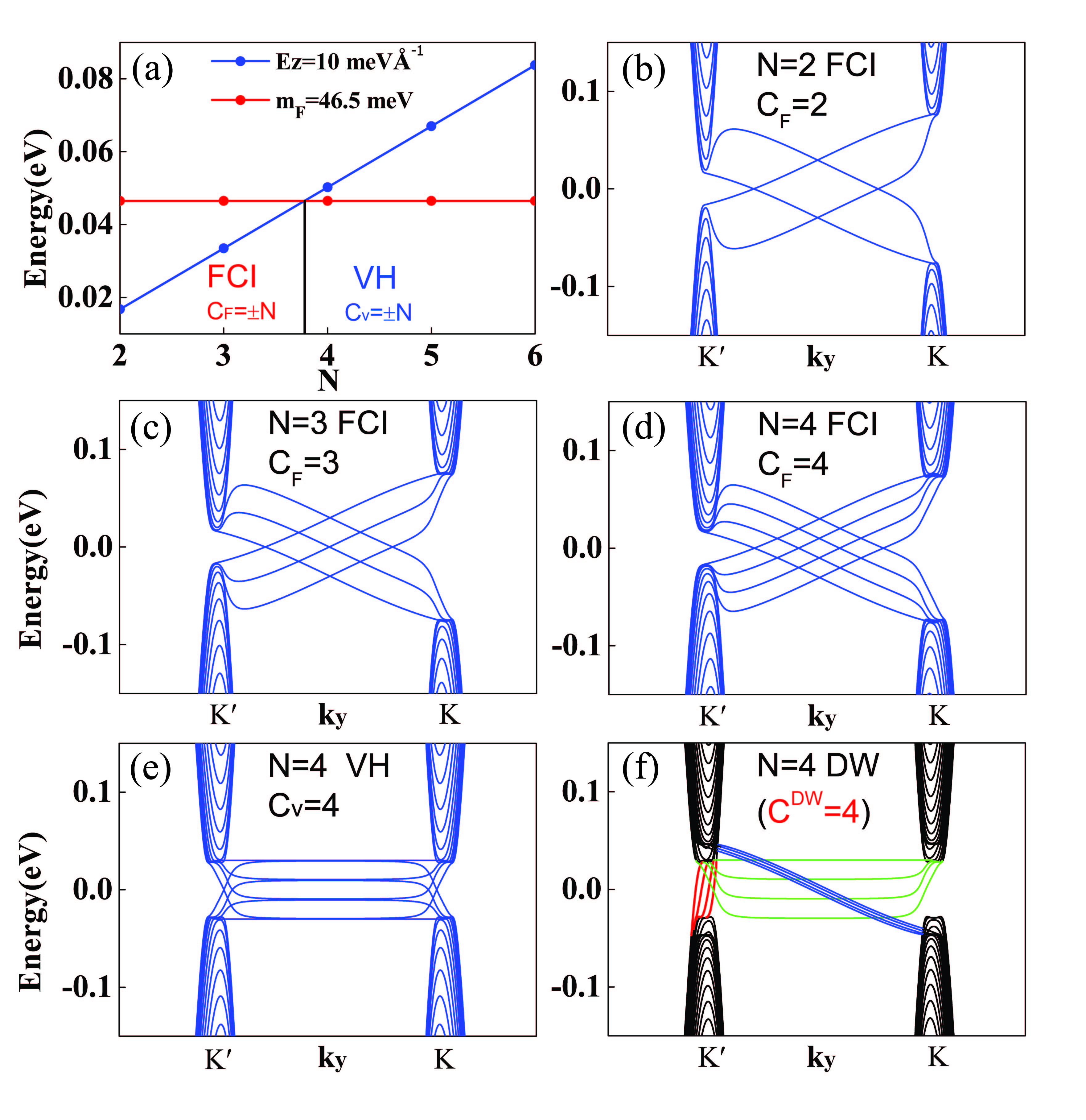}
\caption{Phase diagram and edge states for zigzag ribbons. (a)  Phase diagram with changing the number of the chirally stacked layers.  For $E_z$=10 meV \AA$^{-1}$, $eA_0$=0.15 \AA$^{-1}$, a topological phase transition occurs between $N=3$ and $N=4$. (b)-(d) Showing the chiral edge states of the Floquet Chern insulator (FCI) phases in the $N=2,3,4$ layers with Chern number $C_F=2,3,4$. There are $N(=C_F)$ chiral edge states traverse the bulk gap from one valley to another located at either boundary. (e) Showing the edge states of VH phases in the $N=4$ layers. $N(=C_v)$ flat edge states live at each boundary.  (f) Band structure of a domain wall (DW) made of a FCI and a VH phase. The four red chiral edge states comes from the DW, and only locate at valley $K'$, therefore fully valley-polarized. The bands in black are from the bulk. The four blue (green) edge states in bulk gap originate from the outermost boundary of both FCI (VH) phases. $U$=0.03 eV, $eA_0$=0.15 \AA$^{-1}$, $\xi$=-1 are taken for plotting the band structure.}
\label{Phase_D}
\end{figure}

From the Floquet effective Hamiltonian (Eq.~\ref{Floquet-eff}), by using the Berry curvature formula~\cite{Xiao2010}, the analytical expression for the valence band of the above 2$\times$2 Floquet effective Hamiltonian can be obtained $\varOmega\left(\boldsymbol{p},\tau_{z},\xi\right)=\frac{\tau_{z}}{2} N^2g_{N}^{2}p^{2N-2}\lambda\left(\sqrt{g_{N}^{2}p^{2N}+\lambda^{2}}\right)^{-3}$. After integrating it over the momentum space, one attains the Chern number
\begin{equation}\label{Chern}
C\left(\boldsymbol{p}_{F},\tau_{z},\xi\right)=\frac{N}{2}\tau_{z}\frac{\lambda}{\sqrt{g_{N}^{2}p_{F}^{2N}+\lambda^{2}}},
\end{equation}
where $\boldsymbol{p}_{F}$ is Fermi momentum (the detailed derivation is given in appendix B). If the Fermi level lies in the mass gap, $\boldsymbol{p}_{F}=\boldsymbol{0}$, thus the Chern number has a simple form $C\left(\tau_{z},\xi\right)=\frac{N}{2}\tau_{z}$sgn$\left(U-m_F\xi\tau_{z}\right)$. For the first case, when $U>m_F \left(U<-m_F\right)$, Chern number can be simplified as $C\left(\tau_{z}\right)=\frac{N}{2}\tau_{z}  \left(-\frac{N}{2}\tau_{z}\right)$, thus $C_K=N/2  \left(-N/2\right)$, $C_{K'}=-N/2  \left(N/2\right)$, $C_v=C_K-C_{K'}=N  \left(-N\right)$, and $C_F=C_K+C_{K'}=0  \left(0\right)$, the system belongs to VH phase. For the second case, $-m_F<U<m_F$, the Chern number is written concisely $C\left(\xi\right)=-\frac{N}{2}\xi$, hence $C_K=C_{K'}=-N/2 \left(N/2\right)$, $C_F=-N \left(N\right)$, and $C_v=0 \left(0\right)$ for right (left) CPL, the system realizes the FCIs with high tunable Chern number and a large gap of 93 meV for $eA_0$=0.15 \AA$^{-1}$. Usually, the staggered potential for few-layer graphene can be applied by a gate voltage~\cite{Zou2013}, therefore U is the linear function of layer number $U=\frac{1}{2}E_z\left(N-1\right)d$ with the interlayer distance $d$=3.35 \AA \   and the perpendicular electric field $E_z$  from a gate voltage. Based on the above analysis, we know that Floquet mass favors FCI phases but staggered potential favor VH phases, hence the competition between them will result in topological phase transitions, as shown Fig.~\ref{Phase_D}(a). The staggered potential is linearly increasing and will dominate over the Floquet mass with increasing the number of the layers, resulting in a topological phase transition from a FCI phase to a VH phase with respective topological charge $\pm N$.
 The above analysis, based on a minimal model that can capture the main physics and allows analytical calculations, does not include the warping effect. In fact, when the warping effect is taken into account, we can get the same result qualitatively. (The thorough discussion of the warping effect is given in appendix C).

The above continuum analysis can be verified by the direct calculation of bulk topological invariants and explicit exhibition of the hallmark edge states. In Fig.~\ref{Phase_D}(b)-(e), we plot band structures for chirally stacked and zigzag-terminated $N=2,3,4$-layer graphene ribbons. As for the FCI states, the $N$ chiral edge states connect the conduction band to the valence band in the bulk gap on both boundary of ribbons, as shown in Fig.~\ref{Phase_D}(b)-(d). However, in the VH states (Fig.~\ref{Phase_D}(f)), the $N$ flat edge states connect the conduction (valence) band to the conduction (valence) band on one (the other) boundary.  We also plot the band structure for the DW for $N=4$ layer, as sketched in Fig.~\ref{DW}(a). The four red chiral edge states at $K'$ valley are confined along the DW. The DW localized edge states can be understood by topological charge, as explained in the next section. The four green edge states with flat dispersion come from and are localized at the outermost boundary of the left-hand VH phase. The left four blue chiral edge states in the bulk gap result from and localized at the outermost boundary of the right-hand FCI phase.

\section{Topological domain wall and purely valley-polarized chiral channels}

\begin{figure}[b!]
\includegraphics[width=0.9\columnwidth]{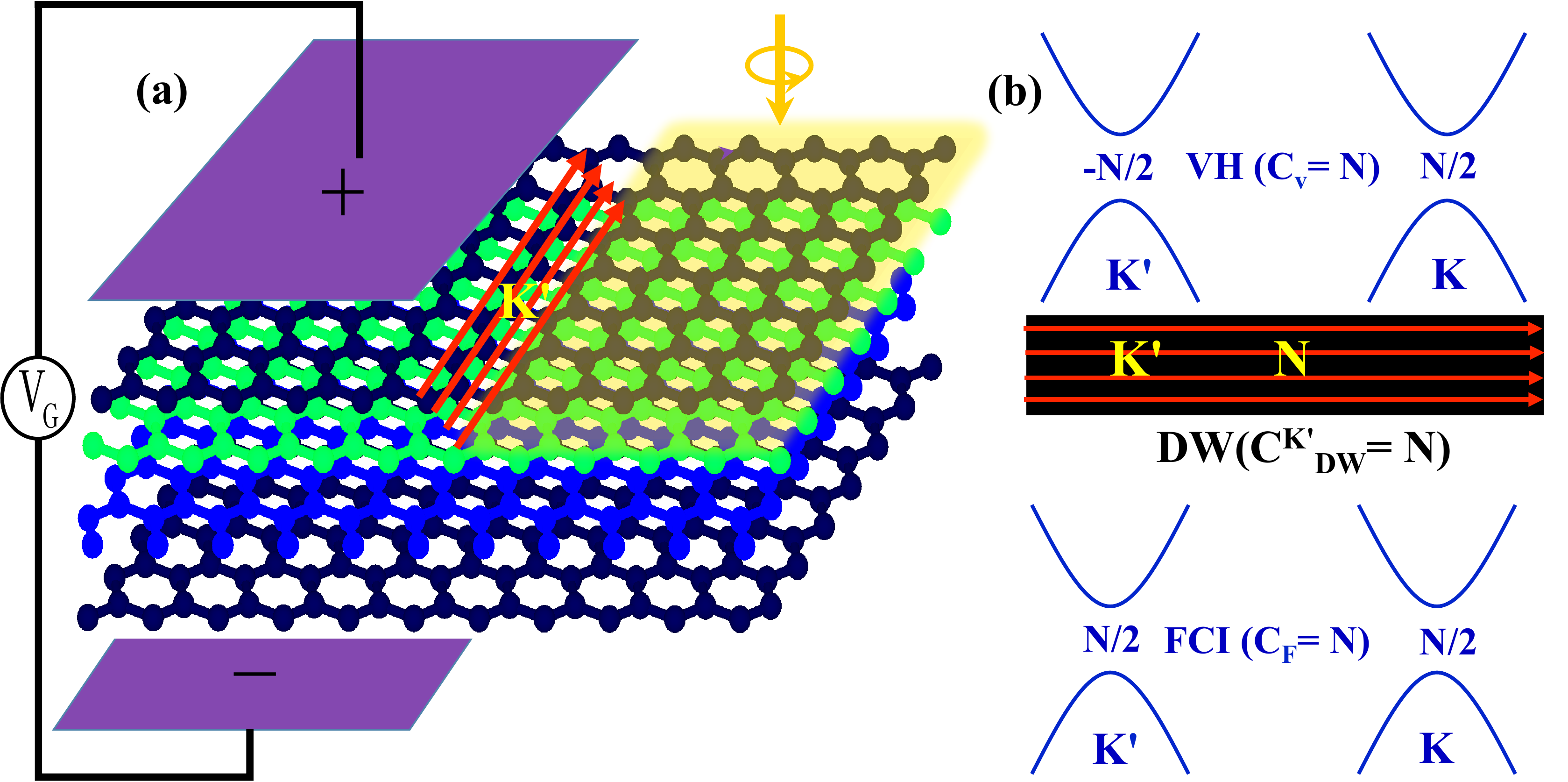}
\caption{(a) Topological domain wall (DW) made of chirally stacked \emph{N}-layer graphene equipped with a gate voltage and a CPL.  (b) The schematic diagram for the valley-projected topological charges distribution for the both sides of the DW in the respective upper and lower panels, and the purely valley-polarized chiral edges with $N$-channel along the DW.}
\label{DW}
\end{figure}

A DW, a type of topological soliton with a discrete symmetry spontaneously broken, can host intriguing edge states~\cite{Martin2008,Semenoff2008,Yao2009,Zhangpnas2013,Vaezi2013,Pan2015,Wang2016}. Using chirally stacked \emph{N}-layer graphene supplemented with a gate voltage and a CPL, a DW is designed, as plotted in Fig.~\ref{DW}(a). On the left part of the topological DW, only a gate voltage is applied, whereas a CPL illuminates its right part.  From the above analysis, one knows that a CPL and a staggered potential can induce respective FCI ($C_F=\pm N$) and VH ($C_v=\pm N$) phases in chirally stacked \emph{N}-layer graphene.
The valley resolved topological charges for the DW read $C^{DW}\left(\tau_{z}\right)=C^{R}\left(\tau_{z}\right)-C^{L}\left(\tau_{z}\right)=\frac{N}{2}\left(-\xi-\tau_{z}sgn\left(U\right)\right)$, where the superscripts DW, L, and R stand for domain wall, and its left and right side.

Here, we use a positive staggered potential ($U>0$) on the left and a left CPL ($\xi=-1$) on the right. As a result, on the two sides emerge a VH phase with $C_v^{L}= N$ and a FCI phase with $C_F^{R}= N$, respectively. The corresponding topological charges for the DW are readily obtained as $C_{K'}^{DW}=N$ and $C_{K}^{DW}=0$. These indicate there are \emph{N} DW edge states propagating in the positive direction with only valley $K'$ index and no channels at valley $K$, as sketched in Fig.~\ref{DW}(b). This analysis is consistent with the above direct calculation of band structure in the zigzag ribbon. The 100$\%$ valley-polarized chiral DW edge states with high channel number but without backscattering will certainly have potential application in valleytronics and low power-consumption devices.

\section{Conclusion}
We have developed a simple and effective route to drive the chirally stacked $N$-layer graphene systems into FCI phases with tunable layer-dependent high Chern number $C_F=\pm N$ and a large gap by using a CPL, which could be verified by using time- and angle-resolved photoemission spectroscopy just as already done in topological insulators~\cite{Wang2013}, and some transport measurement. By applying a gate voltage on the chiral $N$-layers, VH phases with high valley Chern number $C_v=\pm N$ can be induced. Topological phase transitions and anomalous Hall conductivity can be tuned by the size of the gate voltage, and the frequency, field strength, as well as handedness of the CPL. A topological DW is designed to host unidirectionally propagating $100\%$ valley-polarized edges with $N$-channel number, which could have great prospects in application. Our work suggests that the irradiated chirally stacked $N$-layer graphene systems will provide a new playground for future investigations.

\begin{acknowledgements}
The work is supported by the NSF of China (Grant Nos. 11774028, 11734003, and 11574029), the National Key R$\&$D Program of China (Grant No. 2016YFA0300600), the MOST Project of China (Grant No. 2014CB920903), and the Basic Research Funds of Beijing Institute of Technology (Grant No. 2017CX01018).
\end{acknowledgements}

\begin{appendix}
\section{The effective Floquet Hamiltonian}
The Hamiltonian $H\left(\boldsymbol{p},t\right)$ in Eq.~(\ref{time-periodic}) in the main text is time-periodic, it can be expanded as $H\left(\boldsymbol{p},t\right)=\sum_{n}H_{n}\left(\boldsymbol{p}\right)e^{in\omega t}$ with
\begin{equation}
H_{n}=\frac{1}{T}\int_{0}^{T}H\left(\boldsymbol{p},t\right)e^{-in\omega t}dt.
\end{equation}
Hence, we have
\begin{equation}
\begin{split}
&H_{1}=\frac{1}{T}\int_{0}^{T}H_{0}\left(e\boldsymbol{A}\left(t\right)\right)e^{-i\omega t}dt\\
&=\frac{1}{T}\int_{0}^{T}\left(I_{N}\otimes\left(\tau_{z}cos\left(\omega t\right)\sigma_{x}+\xi sin\left(\omega t\right)\sigma_{y}\right)eA_{0}v_{F}\right)e^{-i\omega t}dt\\
&=I_{N}\otimes\left(\tau_{z}\sigma_{x}-i\xi\sigma_{y}\right)\frac{eA_{0}v_{F}}{2},
\end{split}
\end{equation}
\begin{equation}
\begin{split}
&H_{-1}=\frac{1}{T}\int_{0}^{T}H_{0}\left(e\boldsymbol{A}\left(t\right)\right)e^{i\omega t}dt\\
&=\frac{1}{T}\int_{0}^{T}\left(I_{N}\otimes\left(\tau_{z}cos\left(\omega t\right)\sigma_{x}+\xi sin\left(\omega t\right)\sigma_{y}\right)eA_{0}v_{F}\right)e^{i\omega t}dt\\
&=I_{N}\otimes\left(\tau_{z}\sigma_{x}+i\xi\sigma_{y}\right)\frac{eA_{0}v_{F}}{2}.
\end{split}
\end{equation}
$H_{n}=0$ for $\left|n\right|>1$. Here $I_{N}$ is the $N$-order indentity matrix.
In the limit where the driving frequency $\hbar\omega$ is much larger compared to the other energy scales, the effective time independent Hamiltonian is as follow,
\begin{equation}
\begin{split}
H_{F}\left(\boldsymbol{p}\right)&=	H_{0}\left(\boldsymbol{p}\right)+\frac{1}{\hbar\omega}\sum_{n=1}^{\infty}\frac{1}{n}\left[H_{n},H_{-n}\right]+\mathcal{O}\left(\frac{1}{\omega^{2}}\right)\\
&=H_{0}\left(\boldsymbol{p}\right)+I_{N}\otimes\left(-1\right)\frac{\left(eA_{0}v_{F}\right)^{2}}{\hbar\omega}\xi\tau_{z}\sigma_{z}.
\end{split}
\end{equation}

\section{Berry curvature and Chern number}
From Eq.~(\ref{Floquet-eff}) in the main text, we have

\begin{equation}
\begin{split}
H^F_{N,eff}\left(\boldsymbol{p}\right)&=g_{N}\left[\left(\tau_{z}p_{x}+ip_{y}\right)^{N}\sigma_{-}+h.c.\right] +\lambda\sigma_{z}\\
&=g_{N}p^{N}\left[cos(N\phi_{p})\sigma_{x}+sin(N\phi_{p})\sigma_{y}\right] +\lambda\sigma_{z},
\end{split}
\end{equation}
where $cos\phi_{p}=\frac{\tau_{z}p_{x}} {p}$ and $sin\phi_{p}=\frac{p_{y}} {p}$. The eigenvalues and eigenstates of the above Hamilton are given by

\begin{equation}
\begin{split}
E_{c}=\sqrt{h^{2}+\lambda^{2}},\phi_{c}=\frac{1}{\sqrt{h^{2}+(\lambda-E_c)^{2}}}
\left(\begin{array}{c} -he^{-iN\phi_{p}}\\\lambda-E_c\end{array}\right)\\
E_{v}=-\sqrt{h^{2}+\lambda^{2}},\phi_{v}=\frac{1}{\sqrt{h^{2}+(\lambda-E_v)^{2}}}
\left(\begin{array}{c} -he^{-iN\phi_{p}}\\\lambda-E_v\end{array}\right).
\end{split}
\end{equation}
Here $h=\frac{(v_{F}p)^N}{(-t_{\perp})^{N-1}}$. Hence, the Berry curvature for the valence band is
\begin{equation}
\begin{split}
\Omega(\bm p,\tau_{z},\xi)&=-2Im\frac{\left\langle\phi_v\right|v_x\left|\phi_c\right\rangle\left\langle\phi_c\right|v_y\left|\phi_v\right\rangle} {(E_c-E_v)^2}\\
&=\frac{\tau_{z}}{2}\frac{\lambda}{\left(\sqrt{h^2+\lambda^2}\right)^3}\left(\frac{Nh}{p}\right)^2\\
&=\frac{\tau_{z}}{2} N^2g_{N}^{2}p^{2N-2}\frac{\lambda}{\left(\sqrt{g_{N}^{2}p^{2N}+\lambda^{2}}\right)^{3}}.
\end{split}
\end{equation}
Using the Berry curvature, we can obtain the Chern number
\begin{equation}
\begin{split}
C(\tau_{z},\xi)&=\frac{1}{2\pi}\int\Omega(\bm p,\tau_{z},\xi)d^2k \\
&=\frac{N}{2}\tau_{z}\frac{\lambda}{\sqrt{g_{N}^{2}p_{F}^{2N}+\lambda^{2}}}.
\end{split}
\end{equation}

\section{The warping effect}

In the main text, we do not consider the warping effect. In fact, as shows below, the warping effect does not affect our results qualitatively. In the following, we will discuss it detailedly.

When the warping effect is taken into account, the $H_{0}\left(\boldsymbol{p}\right)$ in Eq.~(\ref{time-periodic}) should also include the following terms~\cite{Koshino2009},
$H_{0}\left(\boldsymbol{p}\right)_{i,i\pm3}=\left(v_{3}p_{\pm},0,....,v_{3}p_{\pm}\right)_{2N-3}$ and $H_{0}\left(\boldsymbol{p}\right)_{i,i\pm5}=\left(\frac{\gamma_2}{2},0,....,\frac{\gamma_2}{2}\right)_{2N-5}$, here $p_{\pm}=\tau_{z}p_x\pm ip_y$, $v_{3}=(\sqrt3/2)a\gamma_3/\hbar$. Similarly,  $H_{0}\left(e\boldsymbol{A}\left(t\right)\right)$ in Eq.~(\ref{time-periodic}) should also include an additional term
\begin{equation}
H_{0}\left(e\boldsymbol{A}\left(t\right)\right)_{i,i\pm3}=eA_{0}v_{3}\left(\tau_{z}(\omega t)\pm i\xi sin(\omega t),0,....,\tau_{z}cos(\omega t)\pm i \xi sin(\omega t)\right)_{2N-3}
\end{equation}
besides the term $I_{N}\otimes\left(\tau_{z}cos(\omega t)\sigma_{x}+\xi sin(\omega t)\sigma_{y}\right)eA_{0}v_{F}$. After considered the trigonal warping effect, the full Floquet Hamilton we derived is as follow
\begin{equation}\label{Hwapfull}
\begin{split}
H_{F}\left(\boldsymbol{p}\right)&=
H_{0}\left(\boldsymbol{p}\right)+I_{N}\otimes\left(-1\right)\frac{\left(eA_{0}v_{F}\right)^{2}}{\hbar\omega}\xi\tau_{z}\sigma_{z}\\
&+\xi\tau_{z}\left[I_{1,N-1}\otimes(I_2+\sigma_{z})-I_{2,N}\otimes(I_2-\sigma_{z})\right]\frac{\left(eA_{0}v_{3}\right)^{2}}{2\hbar\omega}\\
&+\mathcal{O}\left(\frac{1}{\omega^{2}}\right),
\end{split}
\end{equation}
where $I_{1,N-1}$ represent the $N\times N$ unit matrix except the last matrix element is zero and $I_{2,N}$ represent the $N\times N$ unit matrix except the first matrix element is zero.
Based on the above full Floquet Hamilton, we can obtain the effective $2\times2$ Floquet Hamiltonian by using the perturbations method~\cite{Koshino2009},

\begin{eqnarray}\label{Hwap2band}
H^F_{N,eff}\left(\boldsymbol{p}\right)&=&\left(\begin{array}{cc}
0 & f(p)\\
f(p)^\dagger & 0
\end{array}\right)+\lambda\sigma_{z},
\end{eqnarray}

\begin{eqnarray}
\begin{split}
f(p)= &\sum\limits_{\left\{n_1,n_2,n_3\right\}}\frac{(n_1+n_2+n_3)!}{n_1 ! n_2 ! n_3 !}\frac{1}{(-\gamma_1)^{n_1+n_2+n_3-1}}\\
&\times(v_F p e^{-i\phi_p})^{n_1}(v_3 p e^{i\phi_p})^{n_2}(\frac{\gamma_2}{2})^{n_3},
\end{split}
\end{eqnarray}
where $\lambda=U-\xi\tau_{z}\frac{1}{\hbar\omega}\left[\left(eA_{0}v_{F}\right)^{2}-\left(eA_{0}v_{3}\right)^{2}\right]$ and the summation is taken over the positive integers which obeys the relation $n_1+2n_2+3n_3=N$.

The introduction of the warping effect makes the analytical calculations impossible. However, we can calculate the Berry curvature and total Chern number numerically by using Eq.~(\ref{Hwapfull}) and Eq.~(\ref{Hwap2band}) for the chirally stacked N-layer graphene. The Chern numbers for $N=2,3,4,5$ layers with specific valley and chirality (the systems with $N>5$ can be calculated in a similar way)  are shown in the Table.~\ref{table1}. The results are perfectly consistent with the analytical expression that without warping effect, i.e., $C(\tau_{z},\xi)=\frac{N}{2}\tau_{z}sgn{(\lambda})$.  In fact, as the warping parameter value $v_3\approx 0.1v_F$~\cite{Koshino2009}, the warping term is much smaller than the leading terms. Hence, the warping term can affect the Berry curvature distribution but not affect the total Chern number. Therefore, the warping effect does not affect our results qualitatively, and the conclusion is valid for the chirally stacked finite N-layer graphene.

\begin{table*}
\caption{\label{table1}The Chern number calculated by Eq.~(\ref{Hwapfull}) and Eq.~(\ref{Hwap2band}). Here, we calculate the systems of $N=2,3,4,5$ layers with $\tau_z=1$ and $\xi=1$ (the systems with $N>5$ can be calculated in a similar way).}
\begin{ruledtabular}
\begin{tabular}{lcr}
&$H_{F}\left(\boldsymbol{p}\right)$&$H^F_{N,eff}$\qquad\qquad\qquad\qquad\\
\hline
$N$ & 2\qquad\qquad 3\qquad\qquad 4\qquad\qquad 5\qquad & 2\qquad\qquad 3\qquad\qquad 4\qquad\qquad 5\qquad\\
\hline
$C(\tau_{z},\xi)$ & -0.996\qquad -1.494\qquad -1.992\qquad -2.493\qquad & -0.999\qquad -1.499\qquad -1.999\qquad -2.499\qquad
\end{tabular}
\end{ruledtabular}
\end{table*}

\end{appendix}

%\bibliography{}

%\end{thebibliography}

\end{document}